\documentclass[12pt]{article}

\newcommand{\eqref}[1]{(\ref{#1})}

\title{Theorem on the Distribution of Short Time Single Particle
Displacements}

\author{R.\ van Zon\footnote{Present address: 
Chemical Physics Theory Group, Department of Chemistry, University of
Toronto, 80 St.\ George St., Toronto, Ontario, Canada M5S 3H6}
\ \ and E.\ G.\ D.\ Cohen\\ 
{\normalsize\it The Rockefeller University, 1230 York Avenue, New
York, NY 10021, USA}}

\date{September 30, 2005}

\begin{document}

\maketitle 

\baselineskip=1.5\baselineskip

\begin{abstract} 
The distribution of the initial very short-time displacements of a
single particle is considered for a class of classical systems with
Gaussian initial velocity distributions and arbitrary initial particle
positions.  A very brief sketch is given of a rather intricate and
lengthy proof that for this class of systems the $n$-th order
cumulants behave as $t^{2n}$ for all $n>2$, rather than as $t^n$.
We also briefly discuss some physical consequences for liquids.
\end{abstract}

\section{Introduction}
This paper is meant to give a pedagogical sketch of the proof of a
theorem for the distribution of initial very short time single
particle displacements in terms of cumulants for a class of classical
systems with smooth potentials, which can be in or out of
equilibrium. This theorem is based on a time expansion around $t=0$.

The theorem has a number of physical
applications\cite{VanZonCohen05c}, including incoherent neutron
scattering in equilibrium
systems\cite{Squires,Rahman64NijboerRahman66}, heterogeneous dynamics
in supercooled glass-formers\cite{Arbeetal02Arbeetal03} and the
recently developed Green's function approach to transport on
picosecond time scales, when the system is far from
equilibrium\cite{Kincaid95,KincaidCohen02aKincaidCohen02bCohenKincaid02}.

In order to formulate the theorems we first introduce the moments
and cumulants of the distribution of single particle displacements.

\section{Moments and Cumulants}

In a classical system of $N$ particles with positions $\mathbf r_i$
and velocities $\mathbf v_i$, we can consider the displacement
$\Delta\mathbf r_i(t)= \mathbf r_i(t)-\mathbf r_i(0)$ of an individual
particle $i$ in a time $t$.  For simplicity, we
will only consider here the displacement $\Delta x_1(t)=\Delta \mathbf
r_1(t)\cdot\hat{\mathbf x}$ of particle 1 in the $x$ direction.

Given the time $t$ and an initial point in phase space, $\Delta
x_1(t)$ is unique. However, we will consider an ensemble of initial
conditions, so that each initial point has a certain probability
associated with it. In fact, the theorem discussed here is applicable
to systems whose initial phase space distribution is of the form
\begin{equation}
  P(\mathbf r^N, \mathbf v^N)=f(\mathbf r^N) \prod^N_{i=1}
  \left(\frac{\beta_im_i}{2\pi}\right)^{3/2}
  \exp\left[-\frac{1}{2}\beta_im_i|{\mathbf{v}}_i-{\mathbf{u}}_i|^2\right]
.
\label{1}
\end{equation}
Here $\mathbf r^N$ and $\mathbf v^N$ represent the collection of the
coordinates $\mathbf r^N \equiv ({\mathbf{r}}_1, {\mathbf{r}}_2,
\ldots, {\mathbf{r}}_N)$ and the velocities $\mathbf v^N \equiv
({\mathbf{v}}_1, {\mathbf{v}}_2, \ldots , {\mathbf{v}}_N)$ of all $N$
particles in the system, respectively,
$\mathbf{u}_i=\langle\mathbf{v}_i\rangle_{t=0}$ for each particle
$i=1,2, \ldots , N$ and $f(\mathbf r^N)$ is a well behaved, i.e.,
normalized, but otherwise arbitrary probability distribution function
of the $\mathbf r^N$ at $t=0$.\footnote{In eq.~(\ref{1}) a
multi-component mixture would have different $m_i$ for particles $i$,
depending on to which component they belong, while $\beta_i$ allows
the particle to belong to a component with a different temperature at
$t=0$ than other particles may have.} We will take ${\mathbf{u}}_i =
0$ in the following for simplicity.

Because of the ensemble distribution of initial phase space points,
$\Delta x_1(t)$ is not fixed but has a probability distribution.  To
characterize this probability distribution function, one may determine
its moments
\begin{equation}
  \mu_n(t) = \langle [\Delta x_1(t)]^n \rangle
\label{2}
\end{equation}
The first moment ($n=1$) is the average of $\Delta x_1(t)$, the second
moment is the average of its square, etc.  In may be noted that for
small times $t$, $\Delta x_1(t)\approx v_{1x} t$, whence one expects
\begin{equation}
  \mu_n(t)\propto t^n.
\label{3}
\end{equation}
The moments may
alternatively be calculated from $\mu_n(t)=\partial^n
F_s(k,t)/\partial(i k)^n|_{k=0}$ using the moment generating function
\begin{eqnarray}
  F_s(k,t) &=& \sum_{n=0}^\infty \mu_n(t) \frac{(ik)^n}{n!} 
\label{4}
\\
           &=& \langle e^{ik\Delta x_1(t)} \rangle.
\label{5}
\end{eqnarray}
Here we used eq.~(\ref{2}) in the last equality. The quantity on the
right hand side of eq.~(\ref{5}) is precisely the self-part of the
intermediate scattering function (the Fourier transform of the Van
Hove self-correlation function) and can be measured through incoherent
neutron scattering\cite{Squires}.  Note that we need in principle all
$\mu_n(t)$ for $n$ from zero to infinity to find $F_s(k,t)$.

Sometimes however, it is known that the first two moments suffice,
namely for an ideal gas and for a perfectly harmonic system, in which
cases $F_s$ and the Van Hove self-correlation
function are Gaussian. This may still be approximately true in less ideal
situations, and in fact, for short times (such as the ones that we are
interested in here) one expects the system to behave almost like an
ideal gas. So it makes sense to try to expand around a Gaussian as a
zeroth order approximation.

In such near-Gaussian cases, the so-called cumulants $\kappa_n$ are a
more convenient set of parameters to work with than the
moments\cite{VanKampen}. The cumulants can be found from the cumulant
generating function $\log F_s(k,t)$, i.e.,
\begin{equation}
  \kappa_n(t)=\frac{\partial^n}{\partial(ik)^n}\log F_s(k, t) \Big|_{k=0}
\label{6}
\end{equation}
so that
\begin{equation}
  F_s(k,t)=\exp\left[\sum^\infty_{n=1} \kappa_n(t)\frac{(ik)^n}{n!}\right]
\label{7}
\end{equation}

Here we will write for convenience the $n$-th cumulant
$\kappa_n(t)\equiv\langle\!\langle \Delta x_1^{[n]}\rangle\!\rangle $,
and stress that $[n]$ is not a power but an index indicating the order
of the derivative of the generating function; for $n=1$ the
superscript will be omitted. Equating the right hand sides of
eqs.~(\ref{4}) and (\ref{7}), one can derive the first few
cumulants in terms of the moments:
\begin{eqnarray}
  \kappa_1(t) & = & \mu_1(t)
\nonumber \\
  \kappa_2(t) & = & \mu_2(t)-\mu^2_1(t)
\nonumber\\
  \kappa_3(t) & = &
  \mu_3(t)-3\mu_2(t)\mu_1(t)+2\mu^3_1(t),
\; \; {\rm{etc.}}
\label{8}
\end{eqnarray}

In general $\kappa_n(t)$ is composed of products of moments of
degree $m$, by dividing its index $n$ in all possible ways into a
sum of integers $m\leq n$, so that their sum $m$ equals $n$.
{}From eq.~(\ref{3}), one would then expect:
\begin{equation}
  \kappa_n(t) \propto t^n.
\label{9}
\end{equation}

\section{The Theorem}

Schofield\cite{Schofield61} and Sears\cite{Sears72} found by
straightforward but complicated calculations for the initial short
time behavior of the $\kappa_n(t)$ the following results  in
equilibrium (where all odd
indexed cumulants vanish):
\begin{equation}
  \kappa_2(t) = \mathcal O(t^2); \kappa_4(t) = \mathcal O(t^8); \kappa_6(t) =
  \mathcal O(t^{12})
\label{10}
\end{equation}
instead of the expected behavior $\kappa_n (t)=\mathcal O(t^n)$
suggested by eq.~(\ref{9}).

Our theorem is a generalization of Schofield and Sears' suggestive
results for the first three non-vanishing cumulants for neutron
scatterings in equilibrium to the general class of systems,
characterized by the initial distribution function~(\ref{1}).

The theorem states that if a) the interparticle and external forces
on the particles are smooth and independent of their
velocities, and b) the velocities are Gaussian distributed
and independent of the initial $\mathbf r^N$ at $t=0$, then
\begin{equation}
  \kappa_n(t)  = 
\left\{\begin{array}{ll}
c_n t^n + \mathcal O(t^{n+1}) & \mbox{ for } n \leq 2
\\
c_n t^{2n} + \mathcal O(t^{2n+1}) & \mbox{ for }   n>2
       \end{array}\right.
\label{11}
\end{equation}
where the $c_n$ depend on the forces but not on $t$.

This result (\ref{11}) implies hidden correlations in the moments
$\mu_n(t)$, which are not eliminated by the simple moment expansion
(\ref{8}) of the cumulants $\kappa_n(t)$.

\section{Sketch of Proof of Theorem}

The proof of the theorem (\ref{11}) proceeds as follows, where
we give only some of the most important steps. The full proof is
presented in ref.~1.

\noindent{\it 1.} We consider systems of $N$ particles $i=1, \ldots,
N$ in $d=3$ with ${\cal{N}} = 3N$ degrees of freedom. It is convenient
for the formulation of the proof to associate with each degree of
freedom of the particles $i$, (generalized) positions $r_i$ and
velocities $v_i$ respectively, but now with $i=1, \ldots,
{\cal{N}}$. Thus $r_{1x}\rightarrow r_1, r_{1y}\rightarrow r_2,
r_{1z}\rightarrow r_3, v_{1x}=v_1, \ldots,$ transforming $\mathbf r^N,
\mathbf v^N$ to $r^{{\cal{N}}}, v^{{\cal{N}}}$. The equations of
motion are then for $i=1, \ldots, {\cal{N}}$:
\begin{equation}
  {\dot{r}}_i =  {v}_i\:;\qquad
  {\dot{v}}_i  =  {F}_i(r^{\cal{N}},t)/m_i=a_i(r^{\cal{N}},t)
\label{12}
\end{equation}

\noindent{\it 2.} 
Expand $\Delta r_1(t)$ in powers of $t$ around $t=0$
\begin{equation}
  \Delta r_1(t)=\sum^\infty_{m=1} \; \frac{t^m}{m!} \;
  \frac{d^m\Delta r_1(t)}{d t^m}\Big|_{t=0}
\label{13}
\end{equation}

\noindent {\it 3.} 
Here the coefficients of $t^m$ are polynomials in $v_1$:
\begin{eqnarray}
  P_1 & \equiv & \frac{d\Delta r_1(t)}{dt}= v_1; 
\nonumber 
\\
  P_2 & \equiv & \frac{d^2\Delta r_1(t)}{dt^2}=a_1(r^{\cal{N}},t)=\mathcal O(1);
\nonumber
\\
  P_3 & \equiv & \frac{d^3\Delta r_1(t)}{dt^3}=\frac{\partial
  a_1}{\partial t} + \sum^{\cal{N}}_{j=1} \frac{\partial
  a_1}{\partial
  r_j}v_j
  = \mathcal O(v^\mathcal N) 
\nonumber 
\\
  P_4 & \equiv & \frac{d^4\Delta r_1(t)}{dt^4} = \ldots +
\sum^{\cal{N}}_{j=1} \; \sum^{\cal{N}}_{k=1} \;
  \frac{\partial ^2 a_1}{\partial r_j \partial r_k} v_j
  v_k=\mathcal O((v^\mathcal N)^2)
  \:\:\: {\rm{etc.}}
\label{14}
\end{eqnarray}

\noindent{\it 4.} 
Therefore, the time expansion of $\Delta r_1(t)$ can
be written in the form:
\begin{equation}
  \Delta r_1(t) = \sum^\infty_{j=1} P_j(v^\mathcal N) t^j
=P_1 (v^\mathcal N) t +
  P_2(v^\mathcal N)t^2+ \ldots + P_n(v^\mathcal N) t^n + \mathcal O(t^{n+1})
\label{15}
\end{equation}
where for $j>1$, $P_j$ is a polynomial of degree $j-2$ in $v^\mathcal N$.

\noindent{\it 5.}
Then the expansion of $\kappa_n(t)$ up to $t^{2n-1}$
can be written in the form:
\begin{eqnarray}
  \kappa_n(t)&\!\!=\!\!&\langle\!\langle \Delta r_1^{[n]} (t)\rangle\!\rangle  
\nonumber 
\\
   &\!\!=\!\!&\!\!\!\!\!\!\!
\mathop{\mathop{\sum_{\{n_j\}} }
_{\sum^n_{j=1} n_j = n }}
_{\sum^n_{j=2} n_j (j-2)<n_1}
\!\!\!\!\!\!\!\!\!
\frac{n!}{n_1! \cdots n_n!} \langle\!\langle v_1^{[n_1]};
  P_2^{[n_2]}(v^\mathcal N); 
\ldots
; P_n^{[n_n]}(v^\mathcal N)\rangle\!\rangle 
  t^{\sum^n_{j=1}n_jj}
\label{16}
\end{eqnarray}
with a correction of $\mathcal O(t^{2n})$.
In this equation, the semicolons on the right hand side of eq.~(\ref{16})
indicate that the $P_m^{[n]}$ inside the $\langle\!\langle
\ldots\rangle\!\rangle $ are not to be multiplied, since they are elements
of a cumulant.

\noindent
{\it 6.} One can prove then that all $n$ terms from $t^n$ to
$t^{2n-1}$ in the sum in eq.~(\ref{16}) vanish, so that only the $\mathcal
O(t^{2n})$ remains and $\kappa_n(t)=\mathcal O(t^{2n})$ (for $n>2$).
The proof of this result is based crucially on the Gaussian properties
of the velocities, so that $\langle v_i^{2n}\rangle=(2n-1)!!\langle
v_i^2 \rangle^n$ obtains.

The theorem can be generalized to the displacement of a single
particle in different directions in a $d$-dimensional space, as well as to
the displacements of different particles, and can also be applied to
multi-component mixtures\cite{VanZonCohen05c}.

\section{Physical Applications} 

The theorem has a number of physical applications, mostly pertaining
to liquids, e.g.  1) It provides a well ordered short time expansion
of the Van Hove self-correlation function relevant for incoherent
neutron scattering in equilibrium systems\cite{Squires}; 2) The
cumulants are connected to the non-Gaussian parameters $\alpha_n$
introduced by Rahman and Nijboer\cite{Rahman64NijboerRahman66} and
which are used as indicators of dynamical heterogeneity in supercooled
glass-formers\cite{Arbeetal02Arbeetal03}; and 3) It provides a well
ordered short time expansion of the Green's functions for far from
equilibrium (mass, momentum and energy) transport on the picosecond
time
scale\cite{Kincaid95,KincaidCohen02aKincaidCohen02bCohenKincaid02}.
For a more extensive account of these applications we refer to ref.~1.

\section*{Acknowledgments}
We would like to thank Prof.\ F.\ Bonetto for useful discussions.
The authors are indebted to the Office of Basic Energy Sciences of
the US Department of Energy under grant number DE-FG-02-88-ER13847.


\begin{thebibliography}{10}
\itemsep 0pt
\parskip 0pt

\bibitem{VanZonCohen05c}
R. van Zon and E.~G.~D. Cohen, cond-mat/0505734; sent to
J. Stat. Phys.

\bibitem{Squires}
G.~L. Squires, {\em Introduction to the Theory of Thermal Neutron
Scattering} (Dover, Mineola, 1996).

\bibitem{Rahman64NijboerRahman66}
A. Rahman, Phys. Rev. {\bf 136}, A405 (1964); B. R. A. Nijboer and
A. Rahman, Physica {\bf 32}, 415 (1966).

\bibitem{Arbeetal02Arbeetal03} 
A. Arbe {\it et al.}, Phys. Rev. Lett. {\bf 89},
245701 (2002); Phys. Rev. E {\bf 67}, 051802 (2003).


\bibitem{Kincaid95}
J.~M. Kincaid, Phys. Rev. Lett. {\bf 74},  2985  (1995).

\bibitem{KincaidCohen02aKincaidCohen02bCohenKincaid02} J.~M. Kincaid
and E.~G.~D. Cohen, Mol. Phys. {\bf 100}, 3005 (2002);
J. Stat. Phys. {\bf 109}, 361 (2002); in: {\em Proc. of the 20th
Symposium on Energy Engineering Sciences} (Argonne National
Laboratory, Argonne, 2002), p.~262.

\bibitem{VanKampen}
N.~G. van Kampen, {\em Stochastic Processes in Physics and Chemistry},
2nd ed.  (North Holland, Amsterdam, 1992).

\bibitem{Schofield61} 
P. Schofield, in {\em Inelastic scattering of neutrons in solids and
liquids} (International Atomic Energy Agency, Vienna, 1961), p.~39.

\bibitem{Sears72}
V.~F. Sears, Phys. Rev. A {\bf 5},  452  (1972).

\end{thebibliography}
\end{document}